\documentclass[12pt]{article}
\usepackage{axodraw}
\begin{document}
\pagestyle{empty}
\begin{flushright}
   hep-ph/0205084\\[-0.1cm]
   September, 2002
\end{flushright}
\begin{center}
{\Large {\bf
New constraint from Electric Dipole Moments  \\
on chargino baryogenesis in MSSM
}}\\[1.cm]
{\large
Darwin Chang$^{a,b}$,
We-Fu Chang$^{a,c}$,
and Wai-Yee Keung$^{a,d}$}\\
{\em $^a$NCTS and Physics Department, National Tsing-Hua University,\\
Hsinchu 30043, Taiwan, R.O.C.}\\
{\em $^b$Theory Group, Lawrence Berkeley Lab, Berkeley, CA94720, USA}\\

{\em $^c$TRIUMF Theory Group, Vancouver, BC V6T 2A3, Canada}\\
{\em $^d$Physics Department,University of Illinois at Chicago, IL
  60607-7059,  USA}\\
\end{center}
\bigskip\bigskip
\centerline{\small \it   ABSTRACT}
\vskip 1cm
A commonly accepted mechanism of generating baryon asymmetry in the
minimal supersymmetric standard model (MSSM) depends on the CP
violating relative phase between the gaugino mass and the Higgsino
$\mu$ term.
The direct  constraint on this phase comes from the limit of electric
dipole moments (EDM)  of various light fermions.
To avoid such a constraint, a scheme which assumes the first two
generation sfermions are very heavy is usually evoked to suppress
the one-loop EDM contributions. We point out that under such a
scheme the most severe constraint may come from a new contribution
to the electric dipole moments of the electron, the neutron or
atoms via the chargino sector at the two-loop level. As a result,
the allowed parameter space for baryogenesis in MSSM is severely
constrained,  independent of masses of the first two generation
sfermions.  \\[0.3cm] PACS numbers: 11.30.Er,11.30.Fs, 12.60.Jv,
98.80.Cq

\newpage
\pagestyle{plain}

While the Standard Model of particle physics continues to accurately
describe a wide array of experimental tests many physicists suspect
that the next generation of a unified field theory will be
supersymmetric.  This supersymmetric theory in its simplest form,
MSSM\cite{mssm}, may help to solve many of the outstanding problems in
the Standard Model.  Two examples of this sort are the
coupling-constant-unification problem and the observed baryon
asymmetry of the universe(BAU).  It is the latter of these two that
will be discussed in this paper.


It has been demonstrated that SM is insufficient in generating large
enough BAU\cite{Farrar:hn}.  Particles lighter in mass but stronger in
coupling are needed to make the electroweak transition more first
order.  Additionally, a new CP violating phase is required to generate
enough BAU.  It is very appealing that MSSM naturally provides a
solution to both requirements\cite{bau}.

The top-quark partner, stop, which is naturally lighter than the
other squarks can make the transition more first order, while
there are plenty of new CP violating phases at our disposal in the
soft SUSY breaking sector.  In particular, it was shown
that the most likely scenario is to make use of the relative phase
between the soft SUSY breaking gaugino mass and the $\mu$ term of
the Higgsino sector\cite{bau}.  In such a case, the BAU is generated
 through the scattering of the charginos with the bubble wall.  The CP
violation is provided by the chargino mixing.  It turns out that
in most parameter space of MSSM, a nearly maximal CP violating
phase is needed to generate enough BAU. One immediate question is
whether or not such  a new source of CP violation is already severely
experimentally constrained.  It is not surprising that the
most severe constraints are provided by the current experimental limits of the
electric dipole moment(EDM) of the electron ($d_e$) and the neutron
($d_n$).

Fortunately the lowest order (one-loop) contributions to various EDM's
through the chargino mixing can be easily suppressed by demanding that
the first two generations of sfermions to be heavier than the third
one\cite{KO,oneloop}.  For example, if one requires these sfermions to
be heavier than $10$ TeV, the one-loop induced EDM's will be safely
small\cite{ref:Abel}.
In fact, such a scenario can even be
generated naturally in a more basic scheme referred to as the more
minimal SUSY model\cite{more}.
However, despite the enlarged parameter space of MSSM, thanks to all
the intricate limits provided by accumulated data from various
collider experiments, there is only a small region of parameters left
within MSSM for such baryogenesis to work\cite{bau}.

In this letter we wish to point out that even if sfermions of the
first two generations are assumed to be very heavy, there are
important contributions to the EDM of the electron at the two-loop
level via the chargino sector that strongly constraint the
chargino sector as the source for BAU in MSSM.  Similar
contributions to quark EDM also exist but the resulting constraint
turns out to be relatively weaker.
While this is not the first time that two-loop contributions are
found to be more important than the one-loop
ones,\cite{ckp,BZ,cky-eEDM,cky-cEDM,oshimo}, this chargino
contribution and its relevance to BAU was never treated fully.

In the case of chargino contributions, the two-loop contribution
is dominant because the one-loop contribution is suppressed when
the sfermions are heavy.  This aspect is similar to those in
Ref.\cite{ckp,oshimo}. In addition, the present case of the large
CP violating phase in the chargino mixing and the light Higgs
scalar, which is necessary to obtain a large baryon asymmetry, is
also the same cause of a large EDM.  Therefore, the resulting
severe EDM constraint is very difficult to avoid in the mechanism of
the chargino baryogensis by tuning parameters.

\section*{The Model and Couplings}
Before we outline the physics of the chargino mixing in
supersymmetric models we will set forth our conventions. We assume
the minimal set of two Higgs doublets. Let the superfield $\Phi_d$
$(Y=-1)$ couple to the $d$-type field, $\Phi_u$ $(Y=1)$ to the
$u$-type  (see Ref.\cite{cky-cEDM} for our convention).
%
%
The chargino fields are combinations of those of the wino
($\omega^+_{L,R}$) and the higgsino ($h^+_{uL, dR}$).
Denote $\psi_L=( \omega^+_L \ ,\ h^+_{uL})^T$,  and
$\overline{\psi_R} = (\overline{\omega^+_R}  \ ,\  \overline{h^+_{dR}})$.
The chargino mass terms,
$ -{\cal L}_M^C = \overline{\psi_R}M_C\psi_L   $
in our convention, becomes
\begin{equation} M_C=
\left(\begin{array}{cc} M_2            &   \sqrt{2}M_W\sin\beta \\
                \sqrt{2}M_W\cos\beta   &   \mu e^{i\phi} \end{array}\right)\ .
\end{equation}
Where  $M_2$ is the $SU_L(2)$ gaugino mass. Note that we choose a CP violating complex
Higgsino mass  $\mu e^{i\phi}$.
The scalar components $H_u, H_d$ of $\Phi_u, \Phi_d$ have real vev's
$v_u/\sqrt{2}, v_d/\sqrt{2}$ respectively, and $\tan\beta=v_u/v_d$.

We use the bi-unitary transformation to obtain the diagonal mass
matrix  $M^D= U' M_C U^\dagger$ with eigenvalues  $m_{\chi_1},m_{\chi_2}$
for the eigenfields $\chi_1, \chi_2$.
SUSY    the The CP violating chargino mixing can contribute to the fermion EDM
through the chargino-sfermion loop.  Detailed analysis of such
contributions can be found in the literature\cite{oneloop}.  As noted
in the introduction, such contributions can be tuned to be small by
making sfermions heavy\cite{ref:Abel} (typically of $10$ TeV or
larger).  Here we are interested in contributions to the EDM of a
fermion that are still important even with very heavy sfermions.  For
this we find that the leading contribution is from diagrams of the
type in Fig.~1.

To evaluate the diagram, we exam gauge couplings of the Higgs
bosons,
$H^0_q= (v_q+\varphi_q)/\sqrt{2}$,
\begin{equation}
{\cal L}_Y =
\hbox{$g\over\sqrt{2}$}\sum_{ij}
\overline{\chi_{iR}} [
U'_{i\omega} U^\dagger_{hj}   \varphi_u^{0*}+
U'_{ih} U^\dagger_{\omega j}  \varphi_d^{0*}    ] \chi_{jL}
 + {\rm H.c.}
\end{equation}
Only the diagonal couplings in the chargino basis are relevant to the
simple diagrams in Fig.\ 1 mediated by an internal photon.
Therefore we define
\begin{equation}
    g^{\varphi_u}_i   \equiv
   g_{iu}^S+ig_{iu}^P =\hbox{$g\over\sqrt{2}$}U'_{i\omega}U_{ih}^*
\ , \quad    g^{\varphi_d}_i   \equiv
   g_{id}^S+ig_{id}^P =\hbox{$g\over\sqrt{2}$}U'_{ih}U_{i\omega}^* \ .
\end{equation}
\begin{center}
\begin{picture}(200,160)(0,0)
\Oval(100,80)(30,50)(0) \Photon(100,110)(100,160){5}{5}
\Text(110,150)[lc]{$\uparrow\ \gamma(k,\mu)$}
\DashArrowLine(20,20)(60,60){5}
\Photon(180,20)(140,60){5}{5}
\Text(30,40)[rc]{$\varphi$}
\Text(50,40)[lc]{$(p)$}
\Line(97,47)(103,53) \Line(97,53)(103,47)
 \Text(105,60)[cc]{$l$}
\Text(140,90)[rc]{$l+q$} \LongArrow(135,102)(140,98)
\Text(60,90)[lc]{$l+p$}
\LongArrow(60,98)(65,102 )   
\LongArrow(85,51 )(80,53)    
\LongArrow(121,53 )(117,51)  
\LongArrow(160,55)(170,45)   
\Text(175,50)[lc]{$\gamma(q,\nu)$}
\ArrowLine(20,20)(0,20) \Text(10,10)[cc]{$f_L$}
\ArrowLine(180,20)(20,20) \Text(100,10)[cc]{$f_R(q)$}
\ArrowLine(200,20)(180,20) \Text(190,10)[cc]{$f_R$}
\end{picture}
\\
Fig.~1. A two-loop diagram of  the EDM of the electron, or quarks.
The chargino runs in the inner loop.
\end{center}
The complex mixing amplitudes are written in terms of
the real couplings $g^S$ and $g^P$.
In the same spirit, the complex neutral Higgs fields are decomposed
into the real and imaginary components $\varphi_q^0=h_q^0+ia_q^0 \
(q=u,d)$.
Note that $h_d^0$ and $h_u^0$ mix in a CP conserving fashion
at the tree-level,  so are $a_u^0$ and $a_d^0$.
\begin{equation}
 \left(\begin{array}{c} h^0 \\ H^0  \end{array}\right)=
  {\cal R}  \left(\begin{array}{c} h_u^0 \\ h_d^0  \end{array}\right)
\ ,\quad
   \left(\begin{array}{c} G^0 \\ A^0  \end{array}\right)=
 {\cal S}  \left(\begin{array}{c} a_u^0 \\ a_d^0  \end{array}\right) \  .
\label{eqn:basis}
\end{equation}
\begin{equation}
    {\cal R}= \left(\begin{array}{cc} \cos\alpha & -\sin\alpha \\
                                 \sin\alpha & \cos\alpha \end{array}\right)
\ ,\quad
       {\cal S}= \left(\begin{array}{cc} \sin\beta & -\cos\beta \\
                           \cos\beta & \sin\beta \end{array}\right)
\ . \label{eqn:angles}\end{equation}
The EDM calculation involves the Higgs boson propagators which are defined
as 
%
\begin{equation}
   \langle\varphi_q\varphi_{q'}^\dagger\rangle_{p^2}
                   = i\sum_\sigma Z_{+,\sigma}^{q,q'}/(p^2-M_\sigma^2) \ ,
   \langle\varphi_q\varphi_{q'}  \rangle_{p^2}
                   = i\sum_\sigma Z_{-,\sigma}^{q,q'}/(p^2-M_\sigma^2) \ .
\end{equation}
%
The $Z$ factors can be shown to be
real at the leading  order with the explicit form,
$$
\begin{array}{ll}
    Z_{\pm,H}^{d,d}= Z_{\pm,h}^{u,u}= \cos^2\alpha\ ,\
& Z_{\pm,G}^{d,d}= Z_{\pm,A}^{u,u}=\pm\cos^2\beta\ ,
\\
    Z_{\pm,h}^{d,d}= Z_{\pm,H}^{u,u}= \sin^2\alpha\ ,\
&  Z_{\pm,A}^{d,d}=Z_{\pm,G}^{u,u}=\pm\sin^2\beta\ ,
\\
Z_{\pm,H}^{u,d}=\hbox{$1\over2$}\sin2\alpha=- Z_{\pm,h}^{u,d}
\ ,
& Z_{\pm,A}^{u,d}=\pm\hbox{$1\over2$}\sin2\beta
=- Z_{\pm,G}^{u,d}
\end{array}$$
$$ Z_{\pm,\sigma}^{d,u} =Z_{\pm,\sigma}^{u,d} \ \hbox{ for }
\sigma=h,H,A,G \ . $$
For completeness, our list includes
the unphysical Goldstone boson  $G^0$, which does  not contribute to EDM.
Other sum rules are
\begin{equation}
 \sum_{\sigma=hHAG}Z_{s,\sigma}^{q,q'}=2 \delta^{q,q'} \delta_{s,+}
\ . \end{equation}
The electron EDM via the Fig.~1 is given by
\begin{equation}
 \left({d_e \over e}\right) = { \alpha \over 16 \pi^3}
                                { g m_e   \over  M_W \cos\beta}
             \sum_{i,q}  {g^P_{i,q} \over m_{\chi_i}}
 \left[  g\left({m_{\chi_i}^2\over M_h^2} \right) Z_{+,h}^{q,d}
       + g\left({m_{\chi_i}^2\over M_H^2} \right) Z_{+,H}^{q,d}
       + f\left({m_{\chi_i}^2\over M_A^2} \right) Z_{+,A}^{q,d}
\right] \ . \label{eqn:bz} \end{equation}
Here the Barr-Zee\cite{BZ} functions are defined as
\begin{equation}
K_n (z)={z\over2} \int_0^1 { y^n\ln{y(1-y)\over z}\over y(1-y)-z}dy \ ,
f(z)=  K_0(z)-2K_1(z)+2K_2(z) \ ,\  g(z)=K_0(z) \ .
\end{equation}

For the EDM of the down quark, we simply use the charge ratio
$1\over3$  to give
$(d_d/e) = {1\over3}(d_e /e)(m_d/ m_e)$.
While for the EDM of the up quark, we need to  replace $Z^{q,d} \to Z^{q,u}$
in  Eq.~(\ref{eqn:bz}) as well as the obvious charge ratio $-{2\over3}$
and replacement of $m_e\to m_u$.
In the Appendix, we offer a more compact analytic form of these
results together with additional details which include the radiative
correction to the Higgs mass in the simplified form suggested in
Ref.\cite{Haber:1996fp}. 

Since the charginos do not couple to the gluon,
there is no chromo-EDM generated\cite{cky-cEDM}.
Note that if one wishes to include  the contribution with the
internal photon  replaced by the $Z$ boson, it is necessary to
include the off-diagonal chargino couplings of the $Z$ and the
Higgs bosons.  We  ignore such  contributions here  because they
are expected to be much smaller than that of the photon which was confirmed in 
previous similar two loop calculations\cite{cky-eEDM}.
In particular, the electron EDM via $Z$ is highly suppressed 
by the small value of the Z vectorial coupling to the electron 
due to the approximate relation $\sin^2\theta_W\approx {1\over4}$.
There are other two loop diagrams with CP violation originated from
the same phase such as the ones with chargino-neutralino loop mediated
$\gamma H^{+} W^{-}$ effective vertex or $\gamma W^{+}W^{-}$ ($W$
EDM) effective vertex.  We do not include them here because these
contributions are expected to be small (by roughly an order of
magnitude) as suggested by previous two loop
calculations\cite{cky-cEDM,oshimo}.  In any case, these additional
diagrams form a separate gauge independent set.

Because the imaginary parts of the off-diagonal entries in $M_C$ are zero in our 
convention, we obtained the following sum rules:
\begin{equation}
\sum_{i} g^P_{i,u} m_{\chi_i}
=-\hbox{$g \over \sqrt{2}$} \mbox{Im} ({U'}^\dagger M_DU)_{\omega h} =0
\ ,\quad
\sum_{i} g^P_{i,d} m_{\chi_i} = 0  \ .
\label{eqn:degen}
\end{equation}
Therefore,
$ g^P_{2,q}=-g^P_{1,q}(m_{\chi_1}/m_{\chi_2})$.
It is easy to see that in case of degenerate masses $m_{\chi_1} =
m_{\chi_2}$, perfect cancellation occurs yielding  zero EDM.

Based upon another fact that the diagonal scalar coupling of
$\bar\chi_i G^0\chi_i$ is zero, we can show that
$ \sin\beta g^P_{i,u} = \cos\beta g^P_{i,d}$.
Therefore, each of the  four CP violating coefficients  $g^P_{i,q}$
can be simply related to one of them, say $g^P_{1,u}$,
which again depends on the fundamental MSSM parameters,
$\tan \beta, \mu e^{i \phi}, M_A^2, M_2$.
The usual SUSY breaking terms include the last two parameters as well as
the trilinear-sfermion-coupling, the $A$ term, which is not relevant in the
our analysis because it does not participate directly in this
particular mechanism of baryogenesis\cite{bau}.
If we replace charginos by stops in the inner-loop,
the effect of the relative phase of $A$ and $\mu$
can contribute to the two-loop EDM as studied in Ref.\cite{ckp}.
The stop-loop effect can be small if $A_t$ is small, if $A_t$ is in
phase with $\mu$, or if the left-handed stop is very heavy but the
right-handed stop is rather light. This last scenario is preferred by
BAU.  Such a large mass gap will suppress stop-mixing and kill the EDM
contribution via the stop-loop.  In addition, it has been concluded by
many groups\cite{bau} that using CP violating mixing of the stop to
generate BAU is much more difficult than using that of the chargino.

\section*{Numerical analysis and baryogenesis}
To our current knowledge, the experimental constraint on the electron
EDM has become very restrictive:
\begin{equation}
|d_e| < 1.6\times 10^{-27}
\hbox{$e$  cm \qquad (90\% C.L. Ref.\cite{eEDM2002})}
\ . \end{equation}
Since the tree-level Higgs mass relation\cite{mssm} predicts a
light Higgs $m_{h^0} < m_Z$ that has already been ruled out by experimental
searches at LEP 2, our analysis has included the leading mass
correction\cite{Haber:1996fp} at the one-loop level.
For completeness, the resulting Higgs mass dependence on $\tan\beta$ in this 
scheme is illustrated in Fig.~2.
Fig.~3 shows the $\tan\beta$ dependence of the predicted value of
the electron EDM from different contributions due to the Higgs bosons, 
$A^0$,  $H^0$ and $h^0$.  
We show the case of maximal CP violation when $\phi=\pi/2$, as required by
baryogenesis\cite{cline}, with masses at the electroweak scale, $M_A=150$
GeV, $M_2=\mu=200$ GeV.  
Note that, in this case, the $h$ contribution dominates until 
about $\tan\beta \approx 3$.  The $H$ contribution becomes dominant for 
$\tan\beta > 5.4$. 
When $\tan\beta$ becomes large, the increase of the
Yukawa coupling of the electron overwhelms the reduction of CP
violation in the chargino sector. This gives the rise of the electron
EDM as $\tan\beta$ increases. The same effect happens to the EDM of
the $d$-quark, but not the $u$-quark.
Fig.~4 shows the electron EDM contour plot versus $M_2$ and $\mu$
for the case $\tan\beta=3$, $M_A=100$ GeV, and $\phi=\pi/2$.
%
%
In the many calculations of BAU in MSSM\cite{bau} the largest
uncertainty seems to come from the calculation of the source term
for the diffusion equations which couples to the left-handed
quarks\cite{cline,murayama}.  Using the latest summary of the
situation in Ref.\cite{cline2} as a reference point, large BAU ($2
\leq \eta_{10} \equiv (n_B - n_{\bar{B}})/ n_\gamma \times 10^{10}
\leq 3$) requires $\tan\beta \leq 3$ with the wall velocity and
the wall width close to their optimal values $v_w \simeq 0.02, l_w
\simeq 6/T$, $\mu \simeq M_2$ and CP phase $\sin \phi$ close to
one.  Note that a smaller $\tan\beta$ gives a larger BAU, however,
it tends to give a small lightest Higgs mass which violates the
LEP II limit unless the left stop is much heavier than 1 TeV.
Using the SUSY parameters in the above range, the numerical
analysis in our figures indicates that the predicted value of the
electron EDM is more than a factor of 5 to 10 bigger than the
experimental limit on the electron EDM in most of the BAU preferred
parameter range.  In fact, if $\sin \phi =1$ and $\tan\beta =3$,
then the parameter space allowed by the electron EDM limit is limited
to a narrow strip with $\mu \simeq M_2$ and $\mu$ has to be as
large as $600$ GeV in order to satisfy this EDM constraint. The range of
values for $\mu$ and $M_2$ (both smaller than $250$ GeV) presented
in Ref.\cite{cline2} are all ruled out.  Unless the numerical
constraint on BAU in Ref.\cite{cline2} is relaxed by an order of
magnitude, it seems to be very difficult for the chargino
mechanism for BAU to be compatible with the electron EDM
constraint.

On the the other hand, for the neutron EDM, our analysis indicates the
current experimental limit in Eq.(12) gives only marginal constraint
on MSSM parameters required for chargino BAU.

With the quark EDM, one uses the quark model to predict the neutron
EDM.
A new limit\cite{nEDMn},  $|d_n| < 6.3 \times 10^{-26} e$ cm
(95\% C.L.), for the neutron EDM has been reported based on the
combination of the recent data of low statistical accuracy and the
earlier measurement\cite{nEDMo}. This combination of the old and
the new results has been criticized in Ref.\cite{nEDMi}.  As shown
in the contour plot of Fig.~5, using the parameters suggested by
the chargino baryogenesis mechanism, our predicted EDM value is around
the size of the more conservative experimental limit:
$|d_n| \stackrel{<}{\sim} 12\times 10^{-26} e$  cm,
recommended in Ref.\cite{nEDMi}.  Due to large theoretical
uncertainties in the relation between the quark EDM and the
neutron EDM, the constraint from neutron EDM on the parameter
space cannot be as important as that from the electron EDM even if
the more stringent limit is used.

Note, however, that the uncertainties in the calculation of non-equilibrium 
electroweak baryogenesis process is far from settle.  
For example, in the latest 
review by the group in Ref\cite{wagner} a small CP violating phase of 
$10^{-2}$ may be sufficient to generate BAU.  
In that case even the larger value 
of $\tan\beta$ is allowed.  
For this purpose, in Fig.~6, we also plot the electron 
EDM for $\tan\beta$ up to $50$. 

\section*{Conclusion}
The baryogenesis in MSSM requires the lightest Higgs boson to be light in
order to get a strong first order phase transition.  It also requires the CP
violating phase in chargino mixing to be large in order to get large enough
BAU.
As we have discussed, both requirements imply the predicted values of
the EDM's of the electron and the neutron to be large.
For $\sin\phi =1$ and $\tan\beta =3$, the current electron EDM
constraint requires $\mu \simeq M_2 \simeq 600$ GeV.  Taking the
uncertainty in the calculations of BAU in the literature into account,
it is probably still premature to claim that this particular mechanism
of baryogenesis is absolutely ruled out, but it is clear that the
precision measurements of the EDM of fermions, especially the electron
EDM, give a tight constraint on the mechanism.

Note added: While the paper is in referee process we receive a manuscript 
(hep-ph/0207277) with calculations that overlap with ours.  Our numerical 
results agree with this later calculation.

WYK is partially supported by a grant from U.S. Department of
Energy (Grant No. DE-FG02-84ER40173). DC is supported by a grant
from National Science Council(NSC) of Republic of China (Taiwan).
We wish to thank H. Haber, H. Murayama, O. Kong, and K. Cheung
for discussions. DC wish to thank theory groups at SLAC and LBL
for hospitality during his visit.  WFC and WYK wish to thank NCTS
of NSC for support.

\section*{Appendix: Higgs Potential with radiative corrections in the MSSM 
and Electric Dipole Moments}

The Higgs potential has the form
\begin{equation}
\begin{array}{rcl}
{\cal  V}
  &=& m_{H_d}^2 |H_d|^2 +  m_{H_u}^2 |H_u|^2
                 +(-m_{12}^2 H_d H_u + \hbox{ H.c. } )  \\
  & &   +\hbox{$1\over8$}(g_1^2+g_2^2) (|H_d|^2-|H_u|^2)^2 
        +{\tau} |H_u|^4   + \cdots  \end{array} \ . 
\end{equation}
At the tree level, SUSY requires the dim=4 coefficient 
$\tau=0$. However, it arises from the
large top-stop-loop correction. 
Denote 
\begin{equation} \begin{array}{lll}
\langle H_d \rangle=V_d \ ,&  \langle H_u \rangle=V_u\ ,&  
V^2\equiv V_d^2+V_u^2   \\
\tan\beta \equiv V_u/V_d \ ,&  m_W^2={1\over2} g_2^2  V^2 \ , &
m_Z^2={1\over2} (g_1^2+g_2^2)  V^2  \ .
\end{array}\end{equation}
We try to derive the mass matrix of the CP-even Higgs bosons, which
correspond to the real part of the complex fields. We 
use superscripts $R,I$ to  abbreviate
the real and imaginary parts.
The first derivatives of the potential are
\begin{equation}\begin{array}{rcl}
 (\partial{\cal V}/\partial H_d^R) &=& 2 m_{H_d}^2 H_d^R - 2m_{12}^2 H_u^R
            +\hbox{$1\over2$} (g_1^2+g_2^2) (|H_d|^2-|H_u|^2) H_d^R \ , \\
 (\partial{\cal V}/\partial H_u^R) &=& 2 m_{H_u}^2 H_u^R - 2m_{12}^2 H_d^R
            -\hbox{$1\over2$} (g_1^2+g_2^2) (|H_d|^2-|H_u|^2) H_u^R
         +4\tau |H_u|^3
\end{array}\ .\end{equation}
The minimization condition can then be written as
\begin{equation}
\begin{array}{r}
 m_{H_d}^2 -m_{12}^2\tan\beta+\hbox{$1\over2$}m_Z^2\cos2\beta=0 \ ,\\
 m_{H_u}^2 -m_{12}^2\cot\beta-\hbox{$1\over2$}m_Z^2\cos2\beta 
                                         +2\tau V^2 \sin\beta=0 \ . 
\end{array} \end{equation}
Continue to obtain the second derivatives,
\begin{equation}
\begin{array}{rl}
(\partial^2{\cal V}/ \partial H_d^{R2}) 
                 &= 2 m_{12}^2 \tan\beta +2 M_Z^2c_\beta^2 \\
 \partial^2{\cal V}/(\partial H_u^R \partial H_d^R) 
                 &=-2 m_{12}^2 -m_Z^2 \sin2\beta \\
(\partial^2{\cal V}/ \partial H_u^{R2}) &
                  =2m_{12}^2 \cot\beta +2 s_\beta^2(M_Z^2 +4\tau V^2) 
\end{array} \ , \end{equation}
\begin{equation}
\begin{array}{rl}
(\partial^2{\cal V}/ \partial H_d^{I2}) & =  2 m_{12}^2 \tan\beta  \\
 \partial^2{\cal V}/(\partial H_u^I \partial H_d^I) &
   = 2 m_{12}^2  \\
(\partial^2{\cal V}/ \partial H_u^{I2}) & =2m_{12}^2 \cot\beta  \end{array} \ .
\end{equation}
The basis defined in Eqs.(\ref{eqn:basis},\ref{eqn:angles})
agrees with  that in Martin's  review\cite{mssm}. 
One can easily  show that $G$ is massless as it is the unphysical
Goldstone boson. 
The mass of the pseudoscalar $A^0$  is
\begin{equation}
 m_{A^0}^2  = 2 m_{12}^2/\sin 2\beta \ ,\quad
   m^2_{H^\pm}= m^2_{A^0} + m_W^2     \ . \end{equation}
The coefficient $m^2_{12}$ corresponds to the non-hermitean quadratic
term in the Higgs potential. 
If $m_{12}^2=0$, the Lagrangian possess a Peccei-Quinn symmetry 
and it quarantees that $M_{A^0}=0$. 
It is practical to express all other masses in terms of $m_{A^0}$. 
From  the second derivatives above, 
the  tree-level mass matrix of the scalar Higgs bosons 
in the basis of $h^0_u,h^0_d$ becomes
\begin{equation} \label{mssmtree}
{\cal M}_0^2=
  \pmatrix{
     m_{A^0}^2\cos^2\beta+m_Z^2\sin^2\beta &
  -( m_{A^0}^2+m_Z^2)\sin\beta\cos\beta \cr
  -( m_{A^0}^2+m_Z^2)\sin\beta\cos\beta &  
     m_{A^0}^2\sin^2\beta+m_Z^2\cos^2\beta\cr} \,,
\end{equation}
where the subscript 0 indicates tree-level quantities.
One can then prove that $(m_{h^0})_0\leq m_Z|\cos2\beta|$.

The leading correction from top-stop loops is
\begin{equation}
{\cal M}_{\rm 1LT}^2 \approx {\cal M}_0^2 +
 T^2 \left( \begin{array}{cc} 1 & 0 \\ 0 & 0\end{array} \right)
\ ,\quad 
 T^2 = 4\tau V^2 s_\beta^2 ={3g^2 m_t^4\over 8\pi^2 m_W^2 \sin^2\beta}
 \ln\left(m_{\tilde t_L} m_{\tilde t_R}/m_t^2\right)    \ . 
\label{topapprox}
\end{equation}
This formula can be found in 
Ref.\cite{Haber:1996fp},
where different schemes of approximation were studied.
As we have uncertaintly from  the SUSY breaking scale, it may be 
overboard to
use the full-fledge 1-loop calculation. We use this leading approximation in
the remaining study.
The CP-even Higgs mass-squared eigenvalues are then given by
\begin{equation}
m^2_{H^0,h^0}=\frac{1}{2}\,\left[{\cal M}^2_{11}+{\cal M}^2_{22}\pm
\sqrt{[{\cal M}^2_{11}-{\cal M}^2_{22}]^2+4({\cal M}^2_{12})^2}\,\right]
\ .\label{massev}
\end{equation}
The mass of $h^0$ has been substantially raised above the tree level
prediction which is lower than the experimental constraint.
The corresponding mixing angle $\alpha$ is given by
\begin{eqnarray} \label{defalpha}
\sin 2\alpha & = & {2{\cal M}^2_{12}\over
\sqrt{[{\cal M}^2_{11}-{\cal M}^2_{22}]^2+4({\cal M}^2_{12})^2}}\,,\nonumber \\
\cos 2\alpha & = & {{\cal M}^2_{22}-{\cal M}^2_{11}\over
\sqrt{[{\cal M}^2_{11}-{\cal M}^2_{22}]^2+4({\cal M}^2_{12})^2}}\,.
\end{eqnarray}
The eigen-masses ($m^2_{H^0} > m^2_{h^0}$) are given by
\begin{equation}
\begin{array}{rl}
 m^2_{H^0}+ m^2_{h^0}     &= m^2_{A^0} + m_Z^2 +T^2   \ ,\\
 (m^2_{H^0}- m^2_{h^0})^2 &=
[(m_{A^0}^2-m_Z^2)\cos 2\beta+T^2]^2 +(m_A^2+m_Z^2)^2\sin^2 2\beta
\ . \end{array}
\end{equation}
In terms of these masses, the mixing angle $\alpha$ 
is determined at tree-level by 
\begin{equation} {\sin 2\alpha\over \sin 2\beta} =
-{m_{A^0}^2 + m_{Z}^2 \over m_{H^0}^2 - m^2_{h^0}} \ ,\qquad\>\>
\cos 2\alpha=
{(m_{Z^0}^2 - m_{A}^2)\cos2\beta-T^2  \over m_{H^0}^2 - m^2_{h^0}}.\>\>\> 
\end{equation}
From the vanishing of diagonal scalar coupling of $\bar \chi G^0\chi $,
we have
$  s_\beta g^P_{i,u} = c_\beta g^P_{i,d} $
for each mass eigenstate $i$.  Therefore 
\begin{equation}\begin{array}{rl}
 \sum_{q}  g^P_{i,q} Z_{+,h}^{q,d}
&=g^P_{i,u} (Z_{+,h}^{u,d} + \tan \beta Z_{+,h}^{d,d}) 
 =g^P_{i,u} (- \hbox{$1\over 2$} \sin 2\alpha + \tan \beta\sin^2\alpha) \\
&=\mbox{$1\over 2$} g^P_{i,u} \tan\beta
    [1 - (m_A^2-4 c_\beta^2 m_A^2 - m_Z^2-T^2)/(m_H^2-m_h^2)]  
\end{array}\end{equation}
\begin{equation}\begin{array}{rl}
 \sum_{q}  g^P_{i,q} Z_{+,H}^{q,d} 
&=g^P_{i,u} (Z_{+,H}^{u,d} + \tan \beta Z_{+,H}^{d,d}) 
 =g^P_{i,u} (\hbox{$1\over 2$} \sin 2\alpha + \tan \beta \cos^2\alpha)  \\
&=\mbox{$1\over 2$} g^P_{i,u} \tan\beta
[1+ (m_A^2-4 c_\beta^2 m_A^2 - m_Z^2-T^2)/(m_H^2-m_h^2)] 
\end{array}\end{equation}
and
\begin{equation}
 \sum_{q}  g^P_{i,q} Z_{+,A}^{q,d}
= g^P_{i,u} (Z_{+,A}^{u,d} + \tan \beta Z_{+,A}^{d,d})  
 = g^P_{i,u} ({1\over 2} \sin 2\beta + \tan \beta \sin^2\beta)  
= g^P_{i,u} \tan\beta \ . \end{equation}
The 2-loop EDM of the electron with the leading 1-loop mass correction becomes 
$$ \left({d_e \over e}\right) 
={ \alpha \over 16 \pi^3} { g m_e   \over 2 M_W \cos\beta}
g^P_{1,u} m_{\chi_1} \tan \beta 
\left[
 \left(1 + { T^2 + M^2_Z + M_A^2(1+2 c_{2\beta}) \over m_H^2-m_h^2}\right)
             {  g(m_{\chi_1}^2/M_h^2) \over m^2_{\chi_1} }
 \right.$$
\begin{equation}
\left.
  +\left(1 - { T^2 + M^2_Z + M_A^2(1+2 c_{2\beta}) \over m_H^2-m_h^2}\right)
             {  g(m_{\chi_1}^2/M_H^2) \over m^2_{\chi_1} }  
  + 2{  f(m_{\chi_1}^2/M_A^2) \over m^2_{\chi_1} } 
  - \Bigl(m_{\chi_1} \rightarrow m_{\chi_2}\Bigr) \right] \ . 
\end{equation}

 
%
\newpage
\begin{center}
\begin{picture}(400,500)(20,0)
\epsffile{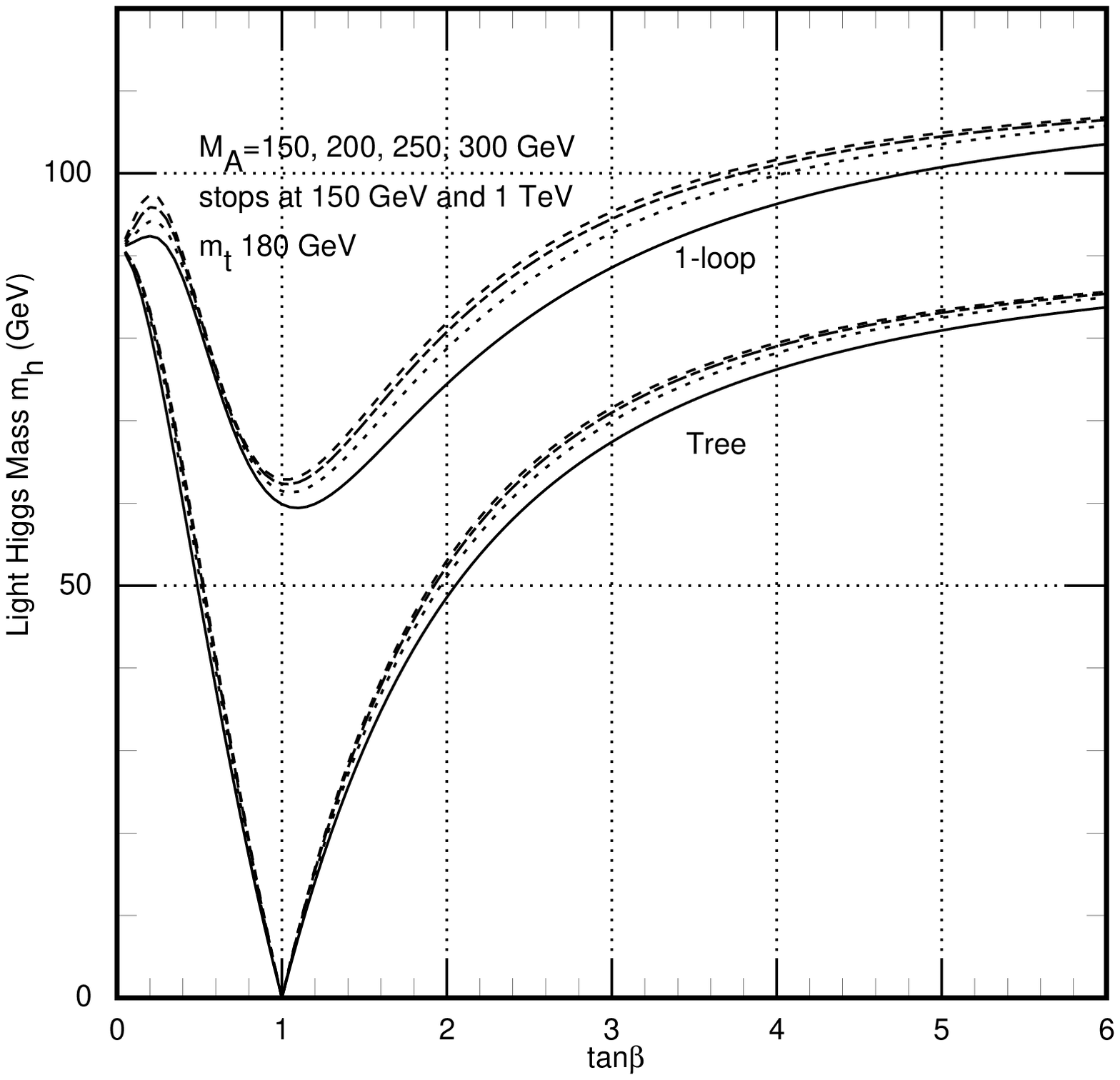}
\end{picture}
\end{center}
\begin{itemize}
\item[Fig.~2]
The mass of the light Higgs boson $h^0$ versus $\tan\beta$.
The lower set of curves corresponds to the tree-level result.
The upper set of curves includes the leading one-loop $(t,\tilde t)$
effect, for $m_{\tilde t_L}=1 $ TeV and  $m_{\tilde t_R}=150 GeV $.
Curves within each set are in the order of cases
$m_A=150,200,250,300$ GeV, from bottom to top.
\end{itemize}
\newpage
\begin{center}
\begin{picture}(400,500)(20,0)
\epsffile{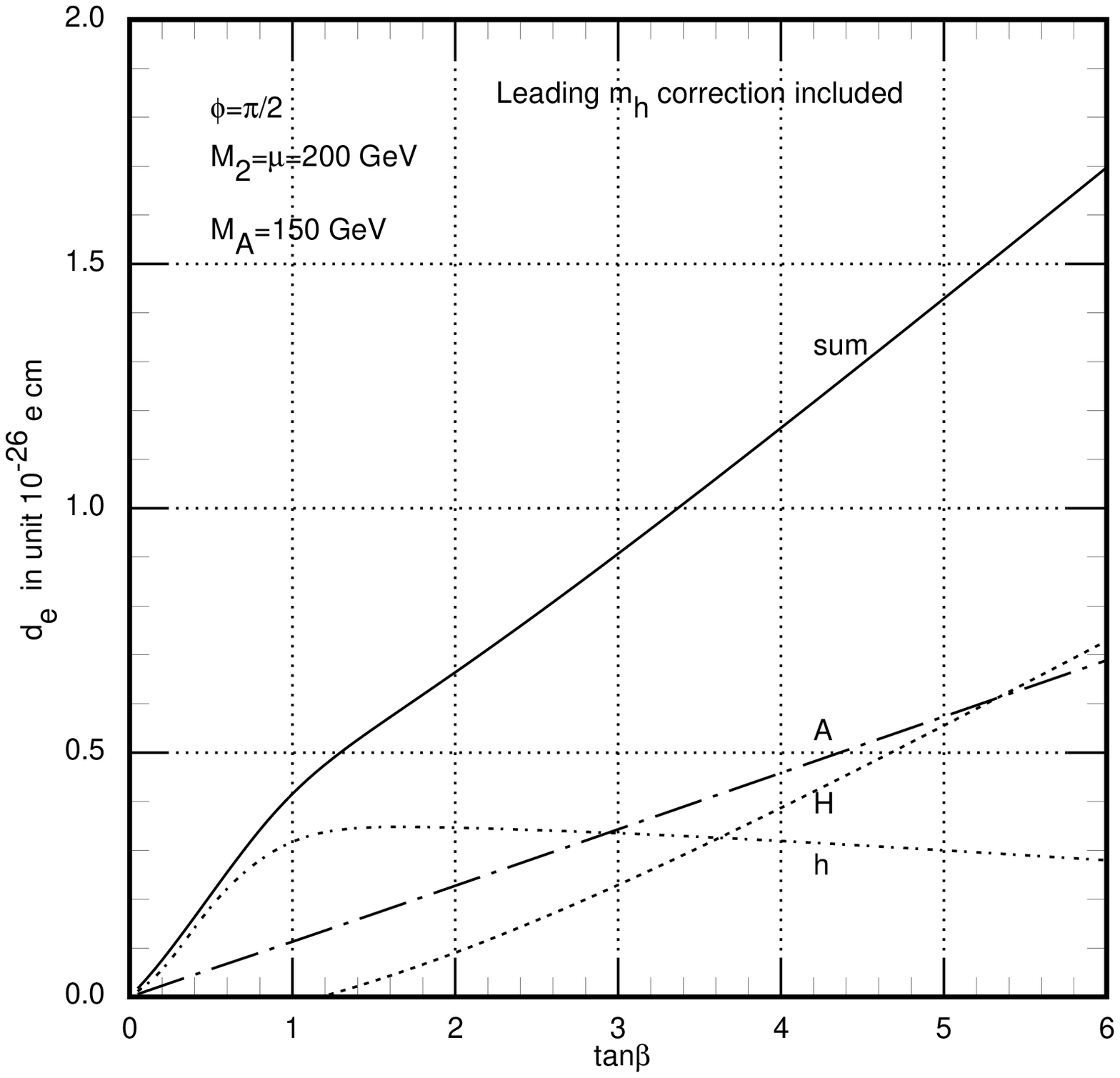}
\end{picture}
\end{center}
\begin{itemize}
\item[Fig.~3]
The predicted value of the
electron EDM versus $\tan\beta$ from different contribtuions due to
the Higgs bosons $h^0, A^0$ and $H^0$,
at the maximal CP violation when $\phi=\pi/2$.
Masses are set at the electroweak scale,
$M_A=150$ GeV, $M_2=\mu=200$ GeV.
\end{itemize}
\newpage
\begin{center}\begin{picture}(400,600)(80,300)
\epsffile{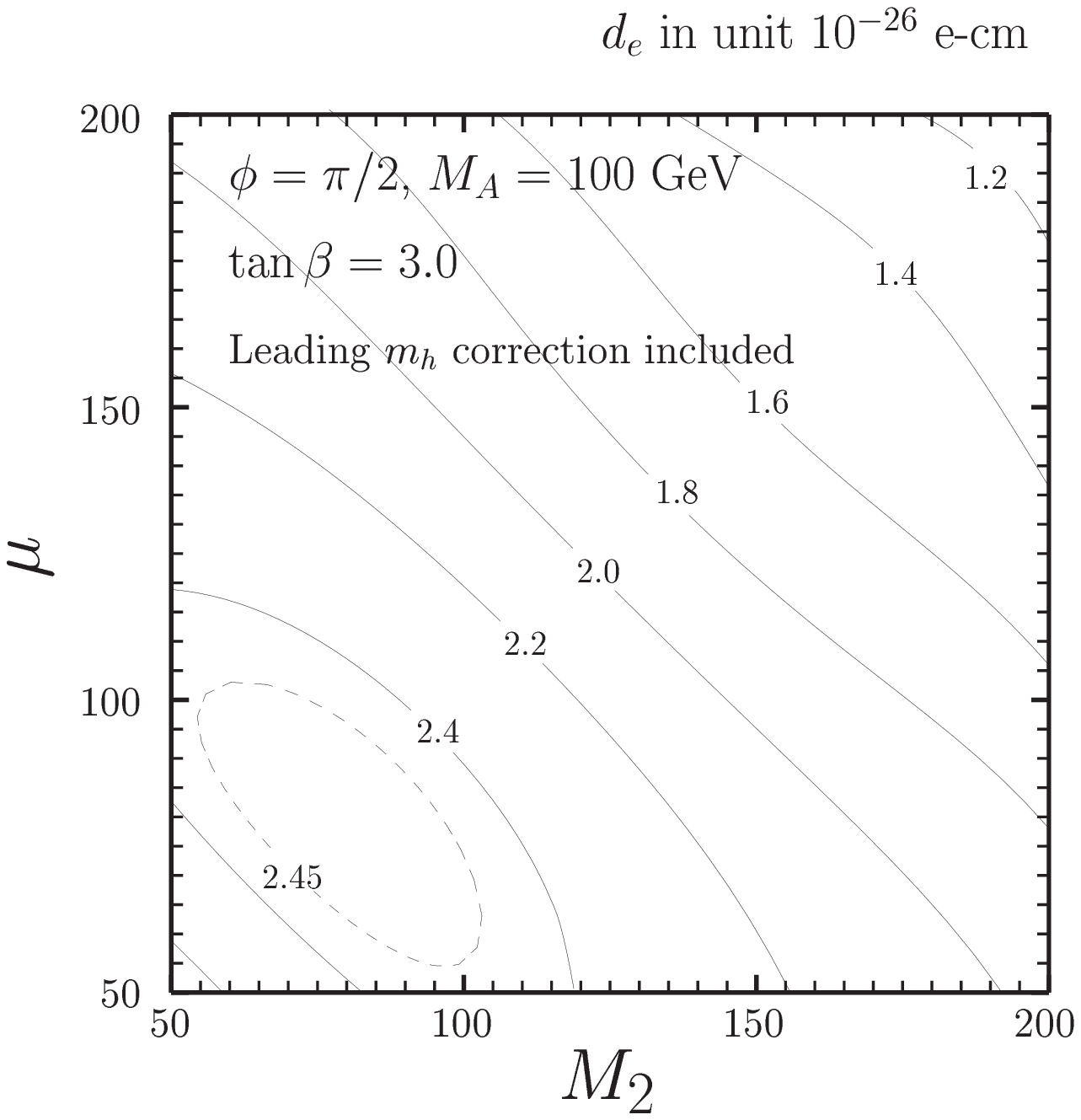}
\end{picture}\end{center}
\begin{itemize}
\item[Fig.~4]
The electron EDM contour plot versus
$M_2$ and $\mu$ for the case $\tan\beta=3$, $M_A=100$ GeV, and
$\phi=\pi/2$.
\end{itemize}
%
\newpage
\begin{center}\begin{picture}(400,600)(80,300)
\epsffile{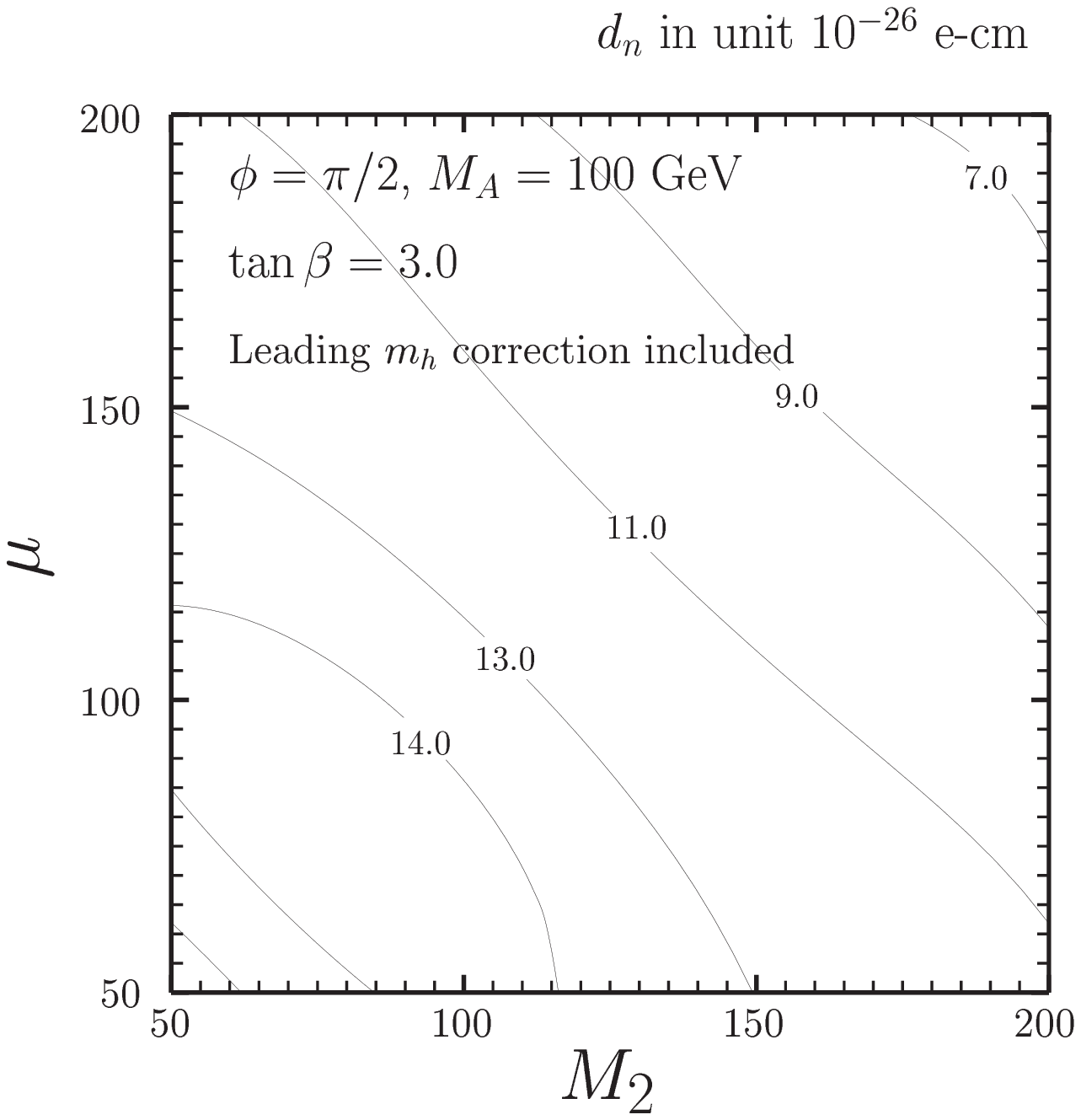}
\end{picture}\end{center}
\begin{itemize}
\item[Fig.~5]
The neutron EDM contour plot versus
$M_2$ and $\mu$ for the case $\tan\beta=3$, $M_A=100$ GeV, and
$\phi=\pi/2$.
\end{itemize}
\newpage
\begin{center}\begin{picture}(400,500)(20,0)
\epsffile{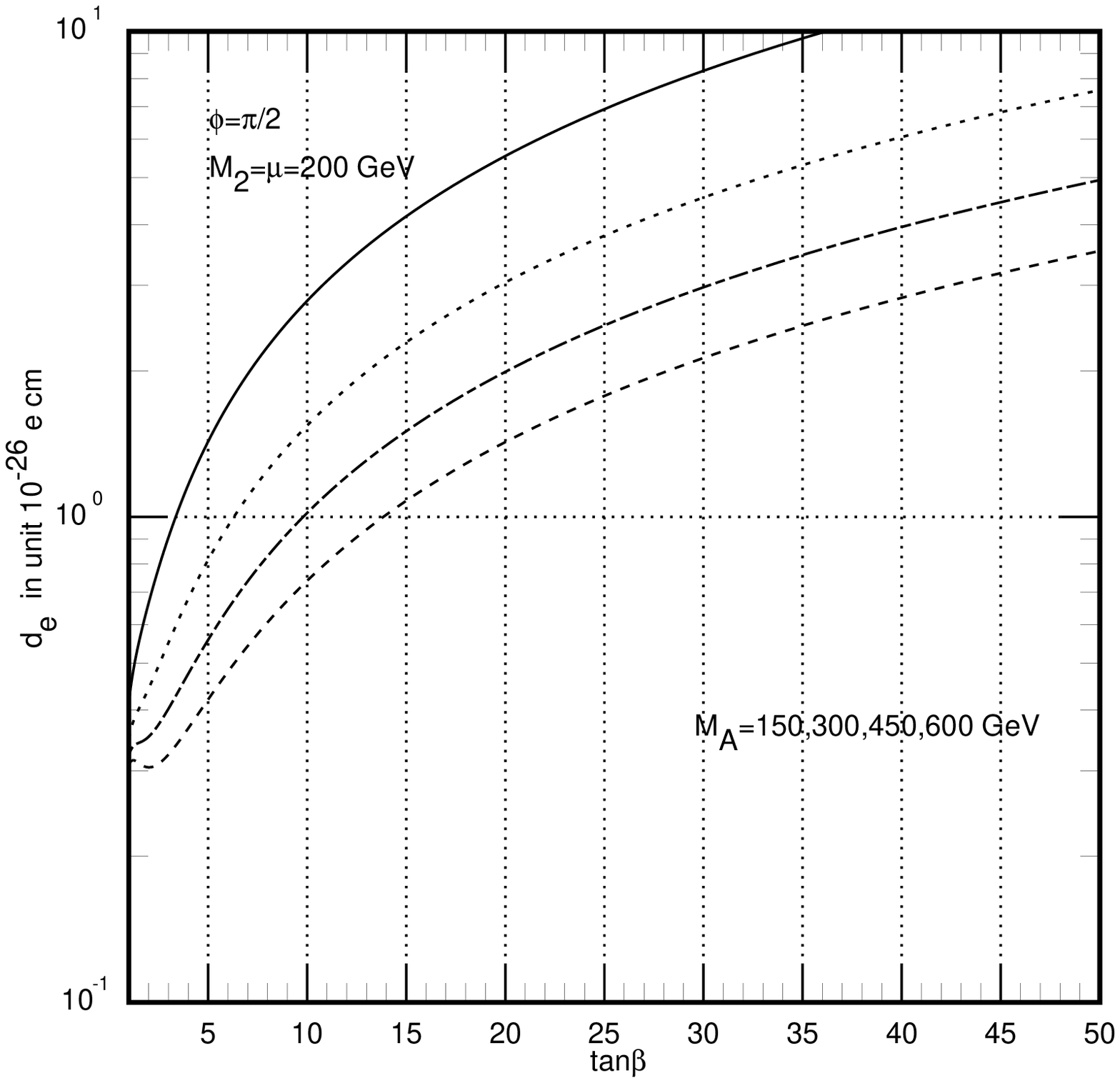}
\end{picture}\end{center}
\begin{itemize}
\item[Fig.~6]
The predicted value of the
electron EDM versus large $\tan\beta$,
at the maximal CP violation when $\phi=\pi/2$.
Masses are set at the electroweak scale,
$M_2=\mu=200$ GeV.
Curves from top to bottom are in the order of cases
$m_A=150,300,450,600$ GeV.
\end{itemize}

\end{document}